\documentclass[twocolumn,amsmath]{aastex702}
\usepackage[T1]{fontenc}
\usepackage{algorithm}
\usepackage{algpseudocode}
\usepackage{wrapfig}

\DeclareSymbolFont{cmletters}{OML}{cmm}{m}{it}
\DeclareMathSymbol{v}{\mathalpha}{cmletters}{"76}

\newcommand{\hammer}{H-AMR}

\definecolor{nick}{HTML}{006400}

\shorttitle{CRUX}
\shortauthors{Liska \& Crozier}

\graphicspath{{./}{figures/}}

\begin{document}

\title{CRUX: A topology-aware load balancer for mesh-based fluid dynamics codes on GPU clusters}

\author[orcid=0000-0003-4475-9345]{M.T.P. Liska}
\affiliation{Center for Relativistic Astrophysics, Georgia Institute of Technology, Howey Physics Bldg, 55 Fifth St NW, Atlanta, GA 30332, USA}
\email[show]{mliska3@gatech.edu}
\correspondingauthor{Matthew Liska}

\author[orcid=0000-0001-9774-2729]{C. Crozier}
\affiliation{H. Milton Stewart School of Industrial \& Systems Engineering, Georgia Institute of Technology, 755 Ferst Dr NW, Atlanta, GA 30332, USA}
\email{CROZIER_EMAIL_REQUIRED}

\begin{abstract}
The rapid growth of computational power has revolutionized numerical simulations, profoundly enhancing our understanding of fluids and plasmas. Computational fluid dynamics (CFD) simulations, which solve partial differential equations governing fluid or plasma motion on discretized grids, have been central to this progress. Recent advances have pushed the resolution and runtime of legacy numerical models to unprecedented levels while enabling newer codes to incorporate increasingly sophisticated physics. However, further scaling of these simulations has become a significant challenge, largely due to the comparatively modest improvements in networking capabilities relative to the rapid growth of floating-point performance in modern GPU-accelerated clusters. In this article, we introduce a novel load-balancing routine \texttt{CRUX} designed to scale efficiently for the most demanding CFD grids in astrophysics. Unlike traditional approaches based on space-filling curves, our method dynamically accounts for computational cost disparities among mesh blocks evolved with different timesteps while minimizing communication overhead. It is also able to take into account heterogeneous hardware. Through an extensive suite of benchmarks featuring up to $5,400$ GPUs on OLCF Frontier and ALCF Aurora, we demonstrate that our load-balancing algorithm outperforms space-filling curve methods across all key metrics, including load uniformity, memory consumption, and communication efficiency, making it a robust solution for next-generation CFD simulations.

\end{abstract}


\section{Introduction}
\label{sec:Intro}
Collisional (and some collisionless) fluids and plasmas are well described by the equations of magnetohydroynamics (MHD). These are a set of partial differential equations (PDEs) consisting out of conservation laws for mass, energy, momentum, and magnetic flux. One popular way of solving these involves discretizing them on a multi-dimensional grid in space and time and solving them using massively parallel computational fluid dynamics (CFD) codes on supercomputers. Such codes allow one to self-consistently simulate demanding problems in physics, astronomy, and engineering while making much fewer assumptions than in semi-analytical calculations. For example, many problems in physics and engineering are inherently turbulent and non-linear, severely complicating the usage of semi-analytical methods, while leveraging the capabilities of MHD. With the advent of petascale and exascale GPU clusters the number of problems that can be addressed by such simulations has dramatically expanded over the last two decades. 

Contemporary MHD simulation codes \citep[e.g.][]{fry00, Gammie2003, Anninos2005, Migone2007, Etienne2015, Porth2017, Hopkins2015, Stone2020, Weinberger2020, Keppens2023, Stone2026, Porth2026} have enabled major advances in plasma astrophysics. Our group, for example, developed the general relativistic magnetohydrodynamic (GRMHD) code \hammer{} \citep{Liska2018A, Liska2020} to simulate plasma accretion onto black holes. These simulations have revealed the complex interplay between the black-hole magnetosphere and the accretion flow. More recently, GRMHD calculations have dramatically improved our understanding of radiatively cooled accretion disks, which can become extremely thin and therefore computationally demanding to resolve \citep{Liska2022, Liska2023, Scepi2023, Zhang2024, Zhang2025, Zhang2026, Fragile2026}. Some of these calculations consequently require the largest GPU-based supercomputers available.

Increasing computational power has also enabled MHD codes to incorporate progressively more complex physics, including radiation \citep[e.g.][]{Sadowski2013, Ryan2015, Foucart2018, White2023} and separate ion and electron thermodynamics \citep{Sadowski2017}. In the near future, we plan to extend \hammer{} with particle-based methods, including Monte Carlo radiation schemes that evolve photon packets alongside the MHD fluid and test-particle methods for evolving non-thermal particle populations \citep[e.g.][]{Bacchini2019, Trent2024, Trent2025}. Unlike the standard MHD equations, whose computational cost is relatively uniform from cell to cell, these additional physical processes can produce large spatial variations in computational expense. For example, radiation source terms may be evolved explicitly in optically thin regions but require substantially more expensive implicit solves at higher optical depths. Such heterogeneous workloads create an increasingly important load-balancing challenge on large GPU clusters.

Typically, MHD codes split up the grid in mesh-blocks with a fixed number of cells per dimension \citep{Berger1989, Balsara2001}. In codes using adaptive mesh refinement (AMR) methods, individual meshblocks can be replaced by meshblocks with a $2$ times smaller cell spacing. Distributing these meshblocks across multiple GPUs on a cluster is challenging since a good load balancing algorithm does not only need to preserve a uniform load, but also needs to minimize the communication cost of ghost cells between meshblocks. These ghost cells are exchanged between neighboring meshblocks and should make minimal use of a cluster's interconnect capability. For example, if neighboring meshblocks exchange ghost cells on a single GPU the cost is less than if ghost cells need to be copied across GPUs. Similarly, it is preferable that ghost cells are exchanged between GPUs on a single node using fast inter-GPU links such as NVIDIA's NVLINK rather than copied between nodes connected to each other via Infinniband switches.

Several algorithms have been proposed that maintain a uniform load balance, some of which account for the communication cost including space-filling curves and graph partiotioners such as \texttt{ParMetis} \citep{Parmetis}. That said, most contemporary astrophysical MHD codes use space-filling curves that are able to translate a 3-dimensional grid to a 1-dimensional curve that can be segmented into uniformly sized pieces which are allocated to individual GPUs. Such space-filling curves naturally colocate neighboring meshblocks on physically close GPUs reducing the cost incurred when exchanging ghost cells between meshblocks. While space-filling curves work extremely well for MHD codes that have a quasi-uniform cost density they are, in our experience, challenging to adapt to more complicated grids that are e.g. evolved on multiple timestepping levels and/or have large inhomogeneity in the cost of each individual meshblocks.

Cost inhomogeneity is not the only issue facing large scale CFD simulations. Most CFD algorithms are floating point intensive, but also require substantial communication of ghost cells between different parts of the grid. Over the past decade, the floating point power of computer hardware has increased much faster than the interconnect bandwidth. This has resulted in many codes becoming limited by the interconnect capability when scaled to a large number of GPUs. To partly address this issue, our group focused on running simulations at higher resolutions with bigger meshblocks, which reduces the relative number of ghost cells. However, this approach makes it difficult to extend the duration of some simulations to physically meaningful timescales. As a recent example, some medium resolution models run in our group typically take months to years to evolve because their grid is too small to scale beyond $\sim 100$ GPUs and the number of timesteps is gigantic. 

In this article we present a new load balancing algorithm that is capable of load balancing the largest GRMHD simulations to date within a few seconds running on up to $5,400$ GPUs on Frontier and Aurora. We first describe the load balancing problem in Section \ref{sec:Problem}, and then introduce our algorithm in Section \ref{sec:algo}. In Section \ref{Sec:HAMR} we describe our implementation of this load balancing algorithm, before presenting our benchmarks in Section \ref{sec:Results} and concluding in Section \ref{sec:Conclusions}.

\section{Load Balancing Problem}
\label{sec:Problem}
\noindent\textbf{Multi-timelevel non-uniform cost grid} State-of-the-art MHD codes typically employ a space-filling curve (SFC), such as a Morton or Hilbert ordering, to partition mesh blocks across GPUs. This approach implicitly assumes that the computational load can be represented by a single scalar weight per block and that balancing the aggregate workload is sufficient to achieve good parallel efficiency. Our setting is more complex. The computational grid is evolved using local adaptive timestepping with multiple time levels $t \in \{0,\dots,T-1\}$. Let $\mathcal{B}$ denote the set of blocks and $\mathcal{B}_g$ the set of blocks on GPU $g \in G$. Each block $b \in B$ has an associated computational cost $c(b)$ and timelevel $\mathrm{l(b)}$. Blocks at lower time levels are updated more frequently by a factor $2^{T-t-1}$ compared to the coarsest timelevel. 

\noindent\textbf{Lock-step evolution:}
Execution proceeds in lock-step: each substep must complete for all blocks $b \in B$ before the next substep begins. Runtime is therefore determined by the maximum per-GPU workload at each substep. Balancing only the aggregate workload defined as the total cost including all timelevels on each GPU like done by traditional space-filling curve approaches,
\begin{equation}
C^{(g)}=\sum_{b \in B_g} c(b) 2^{T-l(b)-1},
\end{equation}
is insufficient, since this can lead to imbalances during the different substeps. 

\begin{figure}
\vspace{-2pt}
\begin{center}
\includegraphics[width=2.55in,trim=0mm 0mm 0mm 0,clip]{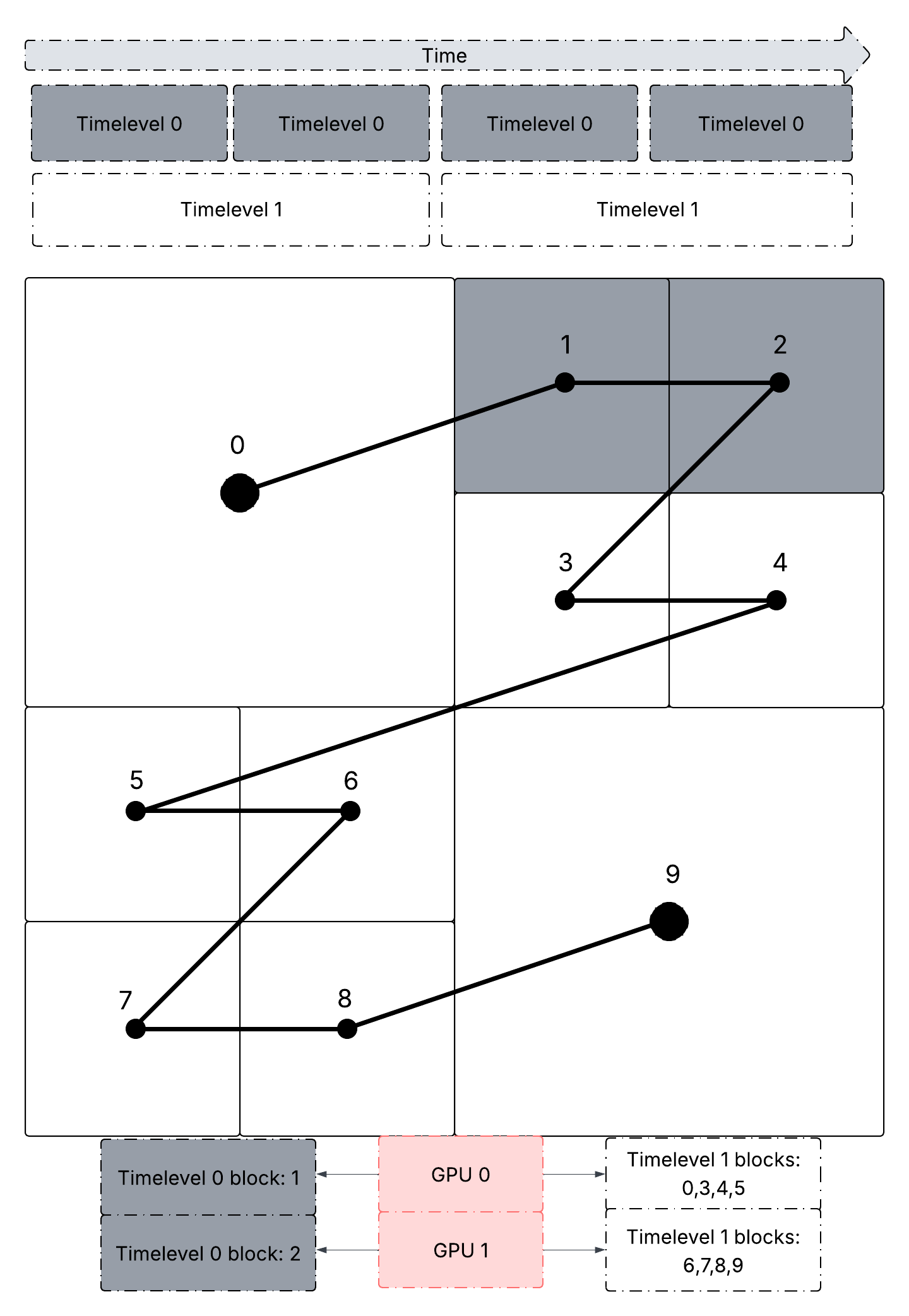}
\vspace{-.05in}
\caption{\footnotesize A split Morton curve is used in stage 0 of our load balancing routine to provide an initial guess for our load balancer. Here the grid features 2 timelevels that are evolved in lock-step. Each of the 2 GPUs gets assigned one timelevel 0 block and 4 timelevel 1 blocks, maintaining load balance at each individual timelevel.
\label{fig:spacefilling}}
\end{center}
\vspace{-4pt}
\end{figure}

For example, consider two GPUs and a workload consisting of two blocks at time level $0$ and eight blocks at time level $1$ (Fig.~\ref{fig:spacefilling}). A global SFC partition may assign two level-0 and two level-1 blocks to GPU 0, and the remaining six level-1 blocks to GPU 1. Over a full synchronization window, the total number of updates is balanced (eight level-1 updates and two level-0 updates). However, the per-substep workload is not balanced. During the first substep, both time levels are active. GPU 0 must process $4$ blocks, while GPU 1 processes $6$. Since execution proceeds in lock-step and finer levels depend on coarser-level data, GPU 0 must wait until GPU 1 completes its $6$ updates before advancing. The effective cost of this substep is therefore equivalent to $6$ updates per GPU, i.e., $12$ effective block updates across the system. In the second substep, only level-1 blocks are active. GPU 1 has no remaining work, while GPU 0 must complete its two level-0 updates. GPU 1 remains idle during this period, resulting in an additional effective cost of $4$ block updates. Consequently, although the nominal workload consists of $8$ block updates, the lock-step schedule results in $12$ effective block updates. In contrast, assigning one level-0 and four level-1 blocks to each GPU equalizes the per-substep critical path and eliminates idle time, yielding strictly better parallel efficiency.

To prevent such load imbalances to occur during the different substeps, we balance the cumulative workload at each timelevel prefix that represents the possible set of blocks evolved during each substep. For each level $t$, we define
\begin{equation}
C_t^{(g)} = \sum_{\substack{b \in B_g \\ \mathrm{level}(b) \le t}} c(b),
\end{equation}
\begin{equation}
N_t^{(g)} = \sum_{\substack{b \in B_g \\ \mathrm{level}(b) \le t}} 1.
\end{equation}
These quantities represent the computational cost and number of blocks active during substeps in which blocks with levels $\mathrm{l(b)} \le t$ are evolved. The effective lock-step execution cost of the grid then becomes the sum of the maximum cost over all GPUs:
\begin{equation}
\mathcal{I}_{C}=\sum_{t=0}^{T-1} 2^{\max(T-t-2,0)} \max_g C_t^{(g)}.
\end{equation}
The weighting factor $2^{\max(T-t-2,0)}$ accounts for the number of substeps that blocks with $\mathrm{l(b)} \leq t$ are evolved on. For example for $T=3$ blocks with $\mathrm{l(b)} \le 0$ are evolved for two substeps (step $0$ and $2$) while blocks with $\mathrm{l(b) }\le 1$ or $\mathrm{l(b)} \le 2$ are evolved during one substep each (step $1$ and $3$).

Load balancing both for the cost and number of blocks has several advantages over only optimizing for the cost. They include: (1) GPUs process multiple blocks in parallel faster than singular blocks. Thus having a fixed number of blocks makes better use of the GPU parallelization that is not capture by e.g. benchmarking singular blocks in serial. (2) Electric fields are synchronized in the middle of each timestep. The part of the calculation before the electric field synchronization has a fixed cost while the part after has a variable cost. By load balancing both the number of meshblocks and the total cost between GPUs we guarantee that both the cost after and before the synchronization step stays uniform. (3) Communication cost is not accounted for in our performance benchmark. By also load balancing the number of blocks we keep the communication cost uniform in the grid.

\medskip
\noindent\textbf{Communication cost model on Frontier:}
To prevent network communication from becoming a performance bottleneck, we minimize the communication cost associated with ghost-cell exchanges between mesh blocks. We adopt a topology-aware weighting scheme based on the Dragonfly interconnect architecture of Frontier, currently the largest DOE open-science GPU system. On Frontier, each node contains four MI250X GPUs, each composed of two multi-chip modules (MCMs). Nodes are connected to a local Ethernet switch, with 16 nodes per switch. Switches within a group are fully interconnected, and groups are connected via global links in a Dragonfly topology. Let $G_{b,t}$ denote the number of ghost-cell bytes exchanged by block $b \in B$ at time level $t \in \{0,\dots,T\}$. If block $b$ resides on GPU $n \in G$ and communicates with a neighboring block on GPU $m \in G$, we assign a topology-dependent penalty weight
\[
w(n,m) =
\begin{cases}
0, & \text{same MCM},\\
1, & \text{same GPU},\\
2, & \text{same node},\\
4, & \text{same switch},\\
8, & \text{same group},\\
16, & \text{different group}.
\end{cases}
\]
This hierarchy reflects increasing communication latency and reduced bandwidth as data traverses progressively larger portions of the interconnect.
The communication cost associated with GPU \(n\) is then defined as
\begin{equation}
\mathcal{C}^{(n)}_{\mathrm{comm}}
=
\sum_{\ell=0}^{L}
\sum_{b\in\mathcal{B}_n}
\sum_{m\in\mathcal{N}(b)}
G_{b,\ell}\, w(n,m),
\end{equation}
where $B_n$ is the set of blocks assigned to GPU $n$, and $\mathcal{N}(b)$ denotes the neighboring blocks of $b$. The global objective minimized by the optimizer is
\begin{equation}
\mathcal{J}_{\mathrm{comm}}
=
\sum_{n}
\mathcal{C}^{(n)}_{\mathrm{comm}},
\end{equation}
which represents the weighted total ghost-cell traffic across all time levels.

\section{Load Balancing Algorithm}
\label{sec:algo}
We refer to our load-balancing strategy as CRUX (Critical-path and Resource-optimized Unified eXchanger). As discussed above, our primary objective is to minimize load imbalance across time levels. Specifically, we aim to reduce the variability of
\[
C_t^{(g)} \quad \text{and} \quad N_t^{(g)} \qquad \forall t \in \{0,\dots,T-1\},
\]
where $C_t^{(g)}$ denotes the cumulative computational cost and $N_t^{(g)}$ the number of active blocks on GPU $g$ at time level $t$. Since execution proceeds in lock-step, runtime is determined by the per-level critical path,
\[
\max_{g} C_t^{(g)} \quad \text{and} \quad \max_{g} N_t^{(g)}.
\]
Accordingly, the primary optimization goal is to minimize these maxima across all GPUs. As a secondary objective, we seek to minimize the communication cost $\mathcal{J}_{\mathrm{comm}}$ without increasing the per-level imbalance. 

To achieve this, we decompose the load balancing procedure into three stages:

\begin{itemize}
\item \textbf{Stage 0:} Construct a baseline partition using a split space-filling curve.
\item \textbf{Stage 1:} Apply a reshuffling algorithm to reduce the per-level imbalance, i.e., to minimize $\max_g C_t^{(g)}$ and $\max_g N_t^{(g)}$ for all $t$.
\item \textbf{Stage 2:} Starting from the balanced configuration obtained in Stage 1, perform communication-aware reshuffling to locally minimize $\mathcal{J}_{\mathrm{comm}}$, subject to the constraint that $\max_g C_t^{(g)}$ and $\max_g N_t^{(g)}$ do not increase for any time level $t$.
\end{itemize}

\noindent This multistage approach is necessary because the minimal achievable imbalance in $C_t^{(g)}$ and $N_t^{(g)}$ is not known a priori. Stage 1 establishes an empirical lower bound on per-level load imbalance. Stage 2 then optimizes communication cost within this fixed load envelope, ensuring that communication improvements do not degrade computational balance.

\subsection{Stage 0: Baseline solution}
Traditional space-filling curve approaches incorporate the number of substeps a block needs to take as part of the cost of each blocks. This approach, however, will not produce homegenous cumulative loads for each timelevel. As we will demonstrate in section ~\ref{sec:Results} this can lead to substantial slow-down when the grid is evolved in lock-step as is the case in most CFD codes. Thus, rather than imposing a single global ordering, we construct an independent Morton (Z-order) curve for each time level \(t \in \{0,\dots,T-1\}\). For a fixed level \(t\), let \(B_t\) be the set of blocks at that level, and let \(\pi_t : \{1,\dots,|B_t|\}\rightarrow B_t\) denote the Morton ordering of \(B_t\). Each block \(b\in B_g\) has cost \(c(b)\). Our goal in Stage~0 is to produce a baseline partition that approximately equalizes the GPU costs $C_t^{(g)}$ for all $t$.

We define the target cost per GPU at level \(t\) as
\begin{equation}
C_{t,\mathrm{target}} \;=\; \frac{1}{G}\sum_{\substack{b \in B_g \\ \mathrm{level}(b) \le t}} c(b).
\end{equation}
We then traverse the ordering \(\pi_t\) sequentially and assign contiguous segments of the curve to GPUs so as to keep each partial sum close to \(C_{t,\mathrm{target}}\). Concretely, we maintain a running sum \(\widehat{C}_t^{(g)}\) for the current GPU \(g\), and append the next block in Morton order to \(g\) while \(\widehat{C}_t^{(g)} < C_{t,\mathrm{target}}\); once this threshold is reached, we advance to GPU \(g+1\) and continue.

When costs are highly non-uniform, the sweep may leave one or more GPUs with no assigned blocks at level \(t\). To avoid empty GPUs, we apply a post-processing step: for any GPU \(g\) with \(B_t^{(g)}=\varnothing\), we reassign one or more blocks from GPUs with \(\widehat{C}_t^{(h)} > C_{t,\mathrm{target}}\) (preferring donor blocks near the segment boundary in Morton order) until all GPUs have nonzero work.

This split-SFC baseline yields an approximately uniform per-level workload, i.e., it reduces the spread of \(C_t^{(g)}\) across GPUs. Moreover, by assigning contiguous Morton segments, it tends to preserve spatial locality and thereby provides a reasonable initial configuration with relatively low ghost-cell communication cost \(\mathcal{J}_{\mathrm{comm}}\).

\subsection{Stage 1: Uniform load}

Stage~1 refines the split space-filling baseline by explicitly minimizing
per-level load imbalance under single-block (1-opt) moves and selected
two-block (2-opt) swaps.  We consider two load metrics: the number of
blocks and the cumulative computational cost per GPU.

\paragraph{Block-count objective.}
For a candidate migration $b: g \rightarrow h$, we define
\begin{equation}
\mathcal{J}_{t}(g,h)
\;=\;
\max\!\bigl(N_t^{(g)},\, N_t^{(h)}\bigr)\, .
\end{equation}
This objective approximates the effective number of substeps contributed
by the pair $(g,h)$ for timelevels smaller than $t$ in a lock-step schedule.

\begin{table*}[ht]
\centering
\caption{Overview of the large-scale GRMHD simulations used to evaluate the load-balancing strategies. For each run we list the number of adaptive timelevels, the total number of meshblocks, the meshblock resolution, the ratio between the most and least expensive blocks in the grid (Max/min Cost), the supercomputing system used, and the total number of GPUs. The MHD runs represent nearly uniform-cost workloads, while the radiative (RAD) simulations exhibit significant computational cost variability due to stiff radiation source terms. The LAT configurations employ local adaptive timestepping with up to five timelevels, introducing additional temporal load imbalance that must be addressed by the load-balancing algorithm in a lock-step compatible way.\label{tab:production_runs}}
\begin{tabular}{lcccccc}
\hline
\hline
Run Name & Timelevels & Meshblocks & Block Size &  Max/min Cost & Cluster & Nr. GPUs \\
\hline
MHD\_LARGE\_NOLAT   & 1  & $96{,}236$   & $48 \times 48 \times 64$ & $1.0$ &  Frontier & $5400$\\
MHD\_LARGE   & 5  & $96{,}236$  &  $48 \times 48 \times 64$ & $1.0$  &  Frontier& $5400$\\
RAD\_LARGE   & 5  & $23{,}185$ &  $48 \times 48 \times 64$  & $4.0$ & Frontier & $2048$\\
RAD\_MID  & 5  & $3{,}335$  &  $56 \times 32 \times 48 $ & $4.2$ &  Aurora & $96$\\
\hline
\end{tabular}
\end{table*}

For each candidate move, we compute $\mathcal{J}_{\rm old}$, apply the move tentatively, recompute $\mathcal{J}_{\rm new}$, and revert the move. A move is admissible if
\[
\mathcal{J}_{t}^{\rm new} \le \mathcal{J}_{t}^{\rm old} .
 \]
Among all admissible moves for GPU $g$, we select the one yielding the largest decrease in $\mathcal{J}$; ties are broken using the communication objective $\mathcal{J}_{\mathrm{comm}}$. If an improving move is found, we permanently migrate the selected block $b^\star$ to GPU $h^\star$ and update all per-level counters. We sweep over all GPUs and repeat until no further single-block move decreases (or tie-improves) $\mathcal{J}$. This produces a locally optimal assignment under 1-opt moves.

\paragraph{Cost-based refinement.}
After block-count balancing, we further refine the partition with respect to cumulative computational cost. For each time level $t$, we define the pairwise cost objective
\begin{equation}
\mathcal{H}_t(g,h)
=
\max\!\bigl(C_t^{(g)},\, C_t^{(h)}\bigr),
\end{equation}
where $C_t^{(g)}$ denotes the cumulative computational cost on GPU $g$ up to time level $t$.

Cost-based updates are restricted to preserve the block-count balance obtained in the previous step. Let $\mathcal{J}_t^{(g)}$ denote the per-level block-count objective. A candidate update is admissible only if
\[
\mathcal{J}_t^{\mathrm{new}} \le \mathcal{J}_t^{\max}
\qquad \forall t \in \{0,\dots,T\},
\]
where $\mathcal{J}_t^{\max}$ is the maximum block-count objective across GPUs after the block-balancing stage.

For each block $b$ currently assigned to GPU $g$, we restrict candidate target GPUs $h$ to those hosting at least one neighbor of $b$. In practice, we consider up to six such neighboring GPUs. If fewer than six distinct neighboring GPUs exist, we additionally sample one GPU uniformly at random from the remaining GPUs to maintain exploration. For each admissible target $h$, we evaluate the cost objective before and after a tentative migration $b:g\rightarrow h$ by computing $\mathcal{H}_t^{\rm old}$ and $\mathcal{H}_t^{\rm new}$ for all affected time levels $t$. The move is accepted only if it reduces the aggregate cost objective while satisfying the block-count constraint above.

In addition to single-block migrations (1-opt), we also consider 2-opt swaps with blocks at the same timelevel. For each candidate GPU $h$ (which by construction hosts neighbors of $b$), we examine blocks $q$ on $h$ that reside at the same time level as $b$. For each such $q$, we tentatively perform the exchange
\[
b:g\rightarrow h,
\qquad
q:h\rightarrow g,
\]
and evaluate the resulting cost objective $\mathcal{H}_t$. The best admissible migration or swap is committed, and the process is repeated until the relative reduction in $\mathcal{H}_t$ falls under a prescribed value after a full sweep through the grid. This balancing stage preserves the previously established block-count balance while reducing cumulative computational imbalance, explicitly accounting for lock-step bottlenecks through the $\max(\cdot,\cdot)$ structure of $\mathcal{J}_t$ and $\mathcal{H}_t$.

\subsection{Stage 2: Communication-aware refinement}
Stage~2 performs a topology-aware reshuffling that minimizes the global communication objective $\mathcal{J}_{\mathrm{comm}}$ while preserving the per-level load balance established in Stage~1.

Recall the global communication cost
\begin{equation}
\mathcal{J}_{\mathrm{comm}}=\sum_{g=0}^{G-1} \mathcal{C}^{(g)}_{\mathrm{comm}},
\end{equation}
which aggregates weighted ghost-cell traffic across all GPUs and time levels.
The objective of Stage~2 is to locally minimize $\mathcal{J}_{\mathrm{comm}}$.

To ensure that communication optimization does not degrade computational balance, candidate updates must preserve the per-level critical path.
Let
\begin{equation}
C_t^{\max} = \max_{g} C_t^{(g)},
\qquad
N_t^{\max} = \max_{g} N_t^{(g)}.
\end{equation}
A tentative update is admissible only if:
\begin{equation}
\max_{g} C_t^{(g),\mathrm{new}} \le C_t^{\max} \qquad \forall t \in \{0,\dots,T-1\},
\end{equation}
\begin{equation}
\max_{g} N_t^{(g),\mathrm{new}} \le N_t^{\max} \qquad \forall t \in \{0,\dots,T-1\}.
\end{equation}
Thus, stage~2 operates within the fixed load envelope obtained after Stage~1 for both the number of blocks and cost per GPU.

Communication reduction is most effective when neighboring blocks are colocated. For a block $b$ currently assigned to GPU $g$, candidate target GPUs $h$ are restricted to those hosting at least one neighbor of $b$. If $b$ has fewer than six distinct neighboring GPUs, we additionally sample one GPU uniformly at random from the remaining GPUs. This restriction reduces the search space to moves with high expected communication benefit. That said we give the user an option to search across all available GPUs. Since we didn't observe any performance benefits in doing so, we do not report further on this alternative design choice.

For each admissible target GPU $h$, we consider both single-block migrations (1-opt) and block exchanges (2-opt). In the 1-opt case, we tentatively migrate $b:g \rightarrow h$ and evaluate $\mathcal{J}_{\mathrm{comm}}^{\mathrm{new}}$. In the 2-opt case, for each block $q$ on GPU $h$ at the same time level as $b$, we tentatively exchange
\[
b:g \rightarrow h,
\qquad
q:h \rightarrow g,
\]
and evaluate the resulting communication objective. Among all admissible candidates satisfying the load-preserving constraint, we select the move or swap yielding the largest reduction in $\mathcal{J}_{\mathrm{comm}}$ and commit it. The procedure is repeated until the relative reduction in $\mathcal{J}_{\mathrm{comm}}$ per sweep falls below a prescribed tolerance. Stage~2 therefore produces a locally optimal configuration under 1-opt and 2-opt updates that minimizes topology-weighted communication cost while provably maintaining the per-substep load balance achieved in Stage~1.

\begin{table*}[ht]
\centering
\caption{Performance comparison between conventional space-filling curve (SFC) load balancing and the proposed CRUX algorithm for several representative GRMHD workloads. All simulations use identical physical setups and hardware configurations; only the load-balancing strategy differs. Columns report the sustained throughput per GPU (zcps/GPU), the maximum GPU memory usage $S_{\max}$, the critical-path imbalance metric $\mathcal{I}$, and the fraction of ghost-zone communication occurring within a single GPU ($f_{\rm GPU}$), between GPUs on the same node ($f_{\rm Node}$), between nodes connected to the same switch ($f_{\rm Switch}$), and across different network groups ($f_{\rm Cluster}$). For uniform-cost MHD runs, CRUX provides a modest improvement in throughput ($\sim$10\%) while maintaining similar communication characteristics. For workloads with significant cost variability (LAT and radiative runs), CRUX reduces the critical-path imbalance $\mathcal{I}$ and improves effective locality, leading to throughput gains of $\sim$5--30\% relative to SFC-based approaches. In particular, for the largest radiative test case (RAD\_LARGE\_LAT), CRUX lowers the imbalance metric by $\sim40\%$ and increases sustained throughput by $\sim30\%$.\label{table:performance}}
\begin{tabular}{llccccccc}
\hline
\hline
Run & Method & zcps/GPU & $S_{\max}(GB)$ & $\mathcal{I_{C}}$ &
$f_{\rm GPU}$ & $f_{\rm Node}$ & $f_{\rm Switch}$ & $f_{\rm Cluster}$\\
\hline

MHD\_LARGE\_NOLAT & SFC-SPLIT
& $0.292 \times 10^8$ & $3.89$ & $21$
& $0.53$ & $0.24$ & $0.08$ & $0.14$ \\

& CRUX
& $0.300 \times 10^8$ & $3.89$ & $21$
& $0.56$ & $0.21$ & $0.08$ & $0.14$ \\

MHD\_LARGE\_LAT & SFC-SPLIT
& $1.28 \times 10^8$ & $4.06$ & $60$
& $0.14$ & $0.36$ & $0.15$ & $0.35$ \\

& CRUX
& $1.41 \times 10^8$ & $3.89$ & $60$
& $0.28$ & $0.25$ & $0.13$ & $0.34$ \\

RAD\_MID\_LAT & SFC
& $2.33 \times 10^6$ & $16.16$ & $1185$
& $0.56$ & $0.24$ & $0.14$ & $0.06$ \\

& SFC-SPLIT
& $4.35 \times 10^6$ & $5.09$ & $449$
& $0.28$ & $0.34$ & $0.13$ & $0.24$ \\

& CRUX
& $4.63 \times 10^6$ & $4.13$ & $407$
& $0.33$ & $0.17$ & $0.22$ & $0.28$ \\

RAD\_LARGE\_LAT & SFC-SPLIT
& $3.80 \times 10^6$ & $3.80$ & $351$
& $0.21$ & $0.39$ & $0.14$ & $0.26$ \\

& CRUX
& $4.96 \times 10^6$ & $3.80$ & $214$
& $0.19$ & $0.02$ & $0.04$ & $0.74$ \\

\hline
\end{tabular}
\end{table*}

\begin{figure*}
\vspace{-0pt}
\begin{center}
\includegraphics[width=\textwidth,trim=0mm 0mm 0mm 0,clip]{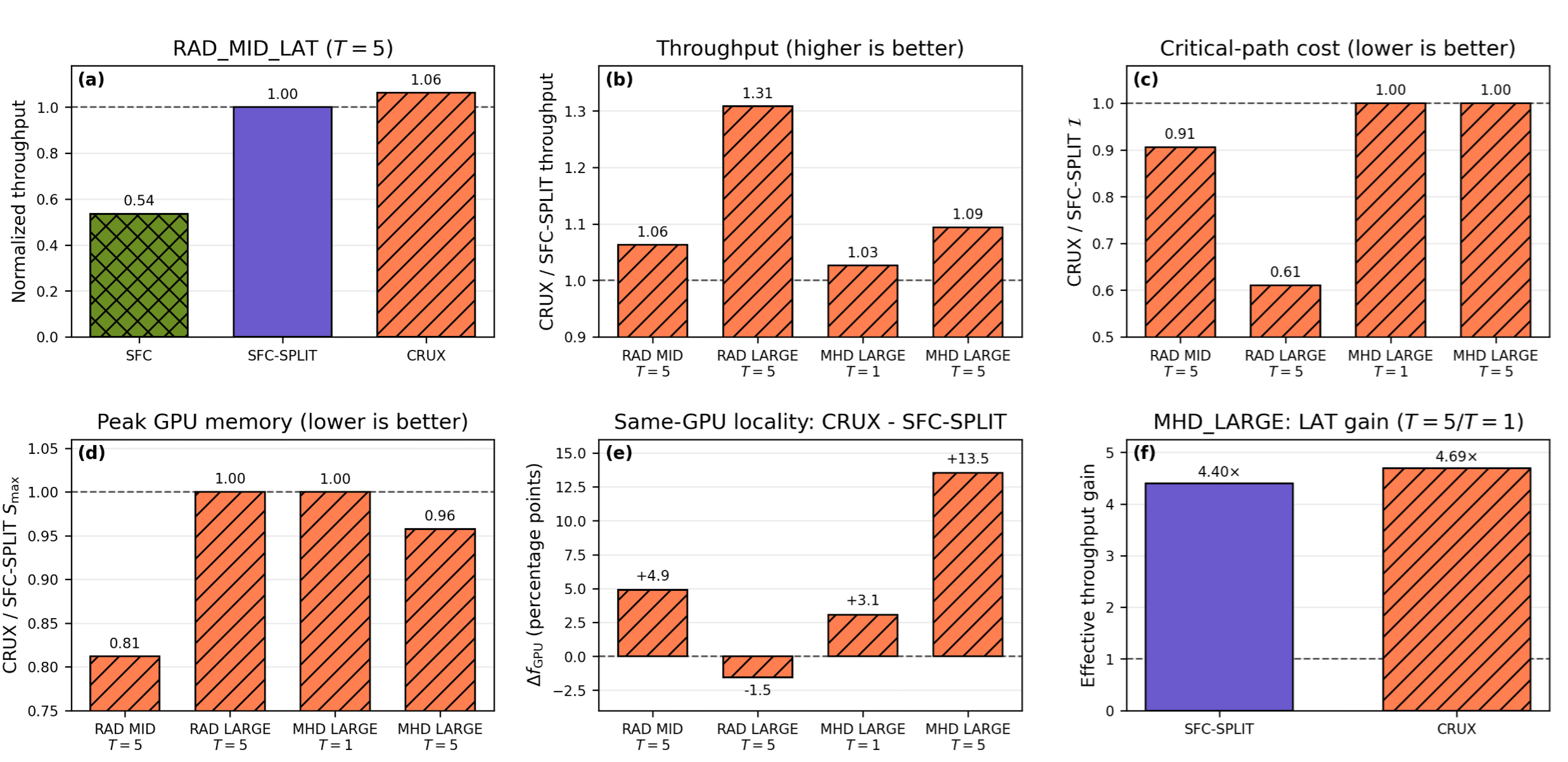}
\vspace{-.0in}
\caption{
Performance impact of the \texttt{CRUX} load-balancing algorithm across representative production GRMHD workloads.
(a) Sustained throughput for the \texttt{RAD\_MID\_LAT} model with five local adaptive timestepping levels ($T=5$), normalized to the SFC-SPLIT result.
(b) Throughput of \texttt{CRUX} relative to SFC-SPLIT for all benchmark models; values above unity indicate a performance improvement.
(c) Critical-path computational cost, normalized to SFC-SPLIT, with smaller values indicating improved load balance and reduced synchronization overhead.
(d) Maximum GPU memory usage relative to SFC-SPLIT, illustrating that the improved computational balance is achieved without increasing peak memory requirements.
(e) Change in the fraction of ghost-zone communication localized within a single GPU, $f_{\rm GPU}^{\rm CRUX}-f_{\rm GPU}^{\rm SFC-SPLIT}$, expressed in percentage points; positive values indicate improved communication locality.
(f) Effective-throughput increase produced by local adaptive timestepping for the \texttt{MHD\_LARGE} model, comparing the $T=5$ calculation with its otherwise equivalent single-timestep ($T=1$) counterpart.
Labels identify both the physical model and maximum number of timestepping levels used in each calculation.
Normalizing quantities within each model removes the large run-to-run variation in absolute computational cost and throughput between MHD and radiative simulations and highlights the effects of the load-balancing strategy itself.
Overall, \texttt{CRUX} provides the largest gains for heterogeneous, multi-timelevel workloads, where it simultaneously reduces the critical path and improves sustained throughput, while remaining close to SFC-SPLIT for the nearly uniform-cost cases.
\label{fig:latbenchmark}}
\end{center}
\vspace{-0pt}
\end{figure*}

\section{Implementation in the \hammer{} GRMHD code}
\label{Sec:HAMR}
We implement our load balancing algorithm into the GPU accelerated general relativistic magnetohydrodynamics (GRMHD) code \hammer{} \citep{Liska2020}. \hammer{} solves the equations of GRMHD on a logically cartesian grid in a covariant format. This means that the user can use any smooth coordinate system without worrying about explicitly adding source terms to the equations as would be necessary in a non-covariant formulation. For example, most simulations with \hammer{} have been performed in spherical coordinates that feature a curved space time around a rapidly spinning black hole. \hammer{} is triple level parallelized with CUDA taking care with most computation on GPUs, OpenMP is used to parallelize gridding and I/O, and MPI is used to facilitate communication of ghost cells between different GPUs (each running a single MPI process). 

Similar to other MHD codes featuring block-based adaptive mesh refinement (AMR) \hammer{} splits up the grid into meshblocks. These meshblocks can have up to an order of magnitude variation in computational cost due to the implicit radiative source terms. In addition, meshblocks are typically evolved at different timestepping levels $t \in T$ through our local adaptive timestepping (LAT) routine. With LAT meshblocks get updated each $2^t$ timesteps rather than every timestep which reduced computational cost by a factor $3-5$ in most usecases. The timestepping level of a meshblock is set by the Courant condition \citep{Courant1928}, which for the purpose of numerical stability limits the timestep to the shortest wave-crossing time of a singular cell part of that respective block. 

Since recently \hammer{} also features a moment-based radiation scheme \citep{Sadowski2013, Mckinney2017} that evolves the energy and momentum densities of a radiation field. This radiation field couples through implicit source terms to the gas modeling physical processes such as emission, absurption and scattering of photons by the gas. The number of Newton-Raphson iterations required for the source terms to converge depend on the properties of the plasma and can vary by up to an order of magnitude accross the grid. Future versions of \hammer{} will feature test particles to model non-thermal electrons and Monte-Carlo radiation to treat multi-frequency radiation. This is expected to make the cost imbalance between meshblocks larger, justifying the development of a load balancer that significantly improves on traditional methods and is scaleable to the largest GPU clusters.

We typically load balance our grid every $\sim 10$ minutes for our production runs. To determine the cost of each meshblock needed by our load balancing scheme, we run a synthetic benchmark before each load balancing step that measures the computational cost of each meshblock in real-time in the presence of stiff radiative source terms. In the absence of radiation, we assume a constant cost which we found to be a good enough approximation. Furthermore, to account for heterogeneuous hardware (such as cluster with multiple types of GPUs), each MPI process can also be assigned a performance factor $p_f$ which is by default set to $p_{f}=1$. The target $N_t^{(g)}$ and $C_t^{(g)}$ are then scaled according to $p_{f}$. As a test, we have load balanced a grid between a NVIDIA Tesla A100 and NVIDIA Gefore RTX 3090 GPU and achieved a $\sim 5 \%$ speedup relative to running on only the A100 GPU, which is $\sim 8$ times faster than the RTX 3090. Since running on heterogeneous architectures is a niche application of our load balancing routine, we won't report on the implementation of the performance factor in our manuscript, but it is included in our open-source code.

\subsection{Parallel Implementation}
While Stage~0 can be efficiently executed using a serial C implementation, stages~1 and 2 of \texttt{CRUX} are dominated by repeated evaluation of candidate 
block migrations and exchanges in a while loop that stops once iterations don't show noticeable improvement. In both stage 1 and 2 the \texttt{CRUX} iterates over meshblocks $b \in B$ and, for each block, evaluates a fixed set of candidate target GPUs $g \in G$. Because the number of candidate GPUs considered per block is bounded and independent of $|G|$, the  computational complexity of a single sweep scales linearly with the  number of meshblocks, i.e., $\mathcal{O}(|B|)$. This structure makes the algorithm naturally amenable to parallelization. Since \texttt{CRUX} has over $3,000$ lines of C code in its current implementation, we only provide a high-level overview of the general parallelization strategy in the CPU implementation.

\subsection{CPU Parallelization}
CRUX is parallelized using OpenMP. The outer loop over blocks is distributed across threads, such that each thread independently evaluates candidate moves for a subset of blocks. For a given block $b$, the thread iterates over its admissible target GPUs (typically neighboring GPUs determined by the communication graph) and records the best admissible move together with the corresponding objective improvement. For simple migrations (1-opt), where a block is moved to a different GPU, the thread stores the proposed destination and tags the block for a potential update in a block-local buffer. If in stage 2 a candidate 1-opt migration indicates that a beneficial 2-opt block exchange may exist, a secondary parallel loop is launched to explore the possible 2-opt swap. In this secondary loop, multiple threads evaluate exchanges between the candidate block and blocks on the target GPU, and subsequently use a parallel reduction algorithm to identify the best admissible swap which are stored in thread-local buffers, before the master thread commits the update to the block-local buffer. 

Because simultaneous updates may conflict (e.g., two blocks attempting to migrate to the same GPU, violating capacity constraints, or invalidating previously computed objective improvements), updates are not applied immediately. Instead, after all threads complete their local evaluation phase, a designated master thread applies the most profitable exchanges stored in block-local buffers sequentially. Each proposed update is revalidated against the current state of the partition to ensure that all constraints remain satisfied and that the objective function still decreases. Only then is the migration committed and the global cost arrays updated. A configurable parameter \texttt{STRIDE} controls how many tentative updates are evaluated before synchronization between threads. Smaller values of \texttt{STRIDE} improve consistency by reducing conflicts between concurrent block exchanged, while larger values increase parallel efficiency. 




\section{Results}
\label{sec:Results}
We evaluate \texttt{CRUX} using large-scale GRMHD simulations performed with \hammer{}, our GPU-accelerated adaptive mesh refinement code \citep{Liska2018A, Liska2022}. These production runs were executed on the OLCF Frontier system (AMD MI250X GPUs) and the ALCF Aurora system (Intel Max GPUs) and represent some of the largest GRMHD calculations performed to date. The simulations include follow-up studies of warped accretion disks \citep{Musoke2022, Kaaz2022, Liska2023} and black hole spectral state transitions \citep{Liska2022, Liska2024}. 

Depending on the physical setup, the computational domain contains between $3\times10^{3}$ and $10^{5}$ meshblocks distributed across $96$–$5{,}400$ GPUs. The effective total resolution spans $N_{\rm cell}\sim10^{8}$–$10^{11}$ cells. We consider both nearly uniform-cost workloads and more challenging configurations that include stiff radiative source terms, producing meshblock cost variations of up to a factor $\sim4$. In addition, some models employ local adaptive timestepping (LAT) with up to five timelevels, while others evolve the grid with a single global timestep. A summary of the simulation configurations and computational characteristics is provided in Table~\ref{tab:production_runs}.

To quantify performance across different timestepping configurations, we measure throughput in terms of both the \emph{effective} and \emph{real} number of zone-cycles per second per GPU,
\begin{equation}
\mathrm{zcps}_{\rm eff}=
\frac{1}{N_{\rm GPU}}
\frac{2^{T-1} N_{T-1} n_{\rm steps} n_{\rm cells}}
{\Delta t_{\rm wall}},
\end{equation}
\begin{equation}
\mathrm{zcps}_{\rm real}=
\frac{1}{N_{\rm GPU}}
\frac{\sum_{t=0}^{T-1}2^{\max(T-t-2,0)}N_t n_{\rm steps} n_{\rm cells}}
{\Delta t_{\rm wall}} .
\end{equation}

Here $n_{\rm cells}$ denotes the number of cells per meshblock, $n_{\rm steps}$ the number of global steps evolved, and $\Delta t_{\rm wall}$ the wall-clock runtime. Each global step consists of $2^{T-1}$ substeps. The quantity $\mathrm{zcps}_{\rm eff}$ represents the throughput that would be obtained if all blocks were updated at every substep and is proportional to the execution speed of the code. On the other hand, $\mathrm{zcps}_{\rm real}$ measures the actual number of zone updates performed during the simulation and is proportional to the amount of work performed on the GPUs. This distinction allows direct comparison between simulations with and without local adaptive timestepping.

\subsection{Performance Benchmarks}
Table~\ref{table:performance} summarizes the performance characteristics of the different load-balancing strategies. In addition to throughput, we report the maximum GPU memory usage $S_{\max}$, the critical-path imbalance metric $\mathcal{I}$, and the fraction of ghost-zone communication occurring within a single GPU ($f_{\rm GPU}$), between GPUs on the same node ($f_{\rm Node}$), between nodes connected to the same switch ($f_{\rm Switch}$), and between different network groups ($f_{\rm Cluster}$). Selected normalized results are visualized in Figure~\ref{fig:latbenchmark}.

Overall, \texttt{CRUX}  consistently improves performance relative to traditional space-filling curve (SFC) partitioning strategies. For nearly uniform-cost grids, the improvement in sustained throughput is modest ($\sim3-5\%$), while for heterogeneous workloads containing stiff radiative source terms the performance gain increases up to $\sim5$–$30\%$.

\subsubsection{Impact of Local Adaptive Timestepping}
Local adaptive timestepping introduces additional complexity for load balancing because all GPUs must still advance substeps in lock-step. Consequently, the effective computational workload varies between substeps and can be difficult to balance using traditional SFC approaches.

Despite these challenges, LAT significantly reduces the overall computational cost of large GRMHD simulations. For example, in Figure \ref{fig:latbenchmark}f the \texttt{MHD\_LARGE\_LAT} run achieves an effective throughput $\sim4.5$ times higher than the corresponding single-timestep run (\texttt{MHD\_LARGE\_NOLAT}), even though the latter exhibits a higher raw update rate $\mathrm{zcps}_{\rm real}$. This behavior reflects the reduced number of zone updates required when LAT is used. The reduction in $\mathrm{zcps}_{\rm real}$ primarily arises from increased communication overhead and reduced GPU saturation during individual substeps, since fewer meshblocks are evolved concurrently.

\subsubsection{Critical Path Cost}
One of the primary advantages of \texttt{CRUX} is its ability to reduce the synchronization bottleneck associated with the lock-step execution model. As shown in Table~\ref{table:performance} and Figure~\ref{fig:latbenchmark}c, the algorithm lowers the critical-path cost metric $\mathcal{I}$ by up to $\sim25\%$ in radiative simulations with heterogeneous workloads. These improvements arise because the algorithm considers block exchanges that are not restricted to the ordering imposed by a space-filling curve, enabling more effective redistribution of computational work across GPUs.

\subsubsection{Memory Usage}
In the current implementation of \hammer{}, memory consumption scales with the number of meshblocks assigned to each GPU. Because \texttt{CRUX} optimizes both computational cost and block counts simultaneously, it produces a more uniform distribution of meshblocks across devices. As shown in Table~\ref{table:performance} and Figure~\ref{fig:latbenchmark}d, the resulting variation in per-GPU memory usage is reduced relative to SFC-based approaches, improving overall utilization of available GPU memory.

\subsubsection{Communication Locality}
\texttt{CRUX} improves communication locality by explicitly accounting for communication costs as defined by the user. Compared to traditional SFC partitioning, a larger fraction of ghost-cell exchanges occur within a single GPU or between GPUs on the same node, while the amount of inter-switch communication is reduced. This improved locality contributes directly to the observed performance gains. In one configuration (\texttt{RAD\_LARGE\_LAT}), the algorithm slightly increases long-distance communication; however, this occurs in exchange for a substantially lower critical-path cost, which ultimately leads to higher sustained throughput.

\section{Conclusions}
\label{sec:Conclusions}
Efficient load balancing is a central challenge for modern GPU-based simulations that combine adaptive mesh refinement, heterogeneous computational workloads, and complex communication patterns. Widely used load-balancing libraries such as \texttt{Zoltan} and \texttt{ParMETIS} typically assume a single scalar weight per mesh block and seek to balance the aggregate workload across processors. However, this formulation becomes inadequate for simulations employing lock-step multi-timelevel timestepping, where the overall runtime is determined by the maximum per-GPU workload during each individual substep rather than by the global workload average. In such cases, load balancing for the total cost could cause the individual substeps to take much longer to complete.

In this work we introduced \texttt{CRUX}, a load-balancing algorithm designed specifically to address these challenges in large-scale CFD simulations that contain multiple timelevels. \texttt{CRUX} explicitly accounts for heterogeneous meshblock costs, lock-step execution across multiple timelevels, and topology-aware communication overhead on modern GPU clusters. Relative to conventional space-filling curve partitioning strategies, \texttt{CRUX} improves sustained performance by $\sim5$–$31\%$ for production-scale MHD and radiative GRMHD simulations. In addition, by jointly optimizing computational cost and meshblock counts, the algorithm reduces the per-GPU memory variation inherent to SFC-based partitions by up to a factor of $\sim3$, leading to significantly more uniform utilization of available hardware resources.

A key advantage of the approach is its modular formulation. The objective function and constraints can be readily adapted to different architectures and application regimes, allowing the load-balancing strategy to reflect the dominant bottlenecks of a given system. For example, our group is developing a particle-based Monte Carlo radiation scheme where we expect to implement additional constraints leading to even larger speed-ups than presented in this article. In the current implementation, the algorithm minimizes total data movement per GPU, which is appropriate when device-level bandwidth limits performance. On systems where inter-switch communication dominates, the communication objective can be reweighted accordingly. Although we have not systematically explored architecture-specific tuning, further performance gains are likely achievable through such adjustments.

Stages 1 and 2 of \texttt{CRUX} are released as open-source software as part of this work and is publicly available at \href{https://bitbucket.org/matthewliska/workspace/projects/CRUX}{Bitbucket}. This enables the user to supply an initial guess to \texttt{CRUX} based on their old partitioning strategy (Stage 0). The current implementation is parallelized using OpenMP and has already been deployed in large-scale \hammer{} production simulations on up to $5{,}400$ GPUs on leadership-class supercomputers. In its present form, the algorithm partitions $\sim10^{5}$ meshblocks across $\sim10^{4}$ GPUs within $\sim10$ seconds on a modern 32-core CPU. A future GPU implementation is expected to extend this capability to grids an order of magnitude larger. These results demonstrate that \texttt{CRUX} provides a practical and scalable load-balancing framework for CFD simulations on modern GPU clusters. More broadly, the approach offers a pathway toward efficiently executing the next generation of exascale astrophysical simulations that combine adaptive mesh refinement and local adaptive timestepping on grids with large inhomogeneities in computational cost.

\begin{acknowledgments}
An award of computer time was provided by the Innovative and Novel Computational Impact on Theory and Experiment (INCITE) program under award AST178. ML was supported by the NASA Hubble Fellowship Program fellowships, NSF grant AST-2407809 and NASA grant 80NSSC22K0817. The authors acknowledge the use of ChatGPT-5 for documenting the respective code for a public release and copy-editing this manuscript. No AI was used for developing the algorithm itself.
\end{acknowledgments}



\bibliography{new.bib,extra.bib}{}

@ARTICLE{Hopkins2015,
       author = {{Hopkins}, Philip F.},
        title = "{A new class of accurate, mesh-free hydrodynamic simulation methods}",
      journal = {\mnras},
         year = 2015,
        month = jun,
       volume = {450},
       number = {1},
        pages = {53-110},
          doi = {10.1093/mnras/stv195},
archivePrefix = {arXiv},
       eprint = {1409.7395},
 primaryClass = {astro-ph.CO},
       adsurl = {https://ui.adsabs.harvard.edu/abs/2015MNRAS.450...53H}
}

@ARTICLE{White2023,
       author = {{White}, Christopher J. and {Mullen}, Patrick D. and {Jiang}, Yan-Fei and {Davis}, Shane W. and {Stone}, James M. and {Morozova}, Viktoriya and {Zhang}, Lizhong},
        title = "{An Extension of the Athena++ Code Framework for Radiation-magnetohydrodynamics in General Relativity Using a Finite-solid-angle Discretization}",
      journal = {\apj},
         year = 2023,
        month = jun,
       volume = {949},
       number = {2},
          eid = {103},
        pages = {103},
          doi = {10.3847/1538-4357/acc8cf},
archivePrefix = {arXiv},
       eprint = {2302.04283},
 primaryClass = {astro-ph.HE},
       adsurl = {https://ui.adsabs.harvard.edu/abs/2023ApJ...949..103W}
}

@ARTICLE{Weinberger2020,
       author = {{Weinberger}, Rainer and {Springel}, Volker and {Pakmor}, R{\"u}diger},
        title = "{The AREPO Public Code Release}",
      journal = {\apjs},
         year = 2020,
        month = jun,
       volume = {248},
       number = {2},
          eid = {32},
        pages = {32},
          doi = {10.3847/1538-4365/ab908c},
archivePrefix = {arXiv},
       eprint = {1909.04667},
 primaryClass = {astro-ph.IM},
       adsurl = {https://ui.adsabs.harvard.edu/abs/2020ApJS..248...32W}
}

@ARTICLE{Stone2020,
       author = {{Stone}, James M. and {Tomida}, Kengo and {White}, Christopher J. and {Felker}, Kyle G.},
        title = "{The Athena++ Adaptive Mesh Refinement Framework: Design and Magnetohydrodynamic Solvers}",
      journal = {\apjs},
         year = 2020,
        month = jul,
       volume = {249},
       number = {1},
          eid = {4},
        pages = {4},
          doi = {10.3847/1538-4365/ab929b},
archivePrefix = {arXiv},
       eprint = {2005.06651},
 primaryClass = {astro-ph.IM},
       adsurl = {https://ui.adsabs.harvard.edu/abs/2020ApJS..249....4S}
}

@inproceedings{Parmetis,
  author    = {George Karypis and Vipin Kumar},
  title     = {A Parallel Algorithm for Multilevel Graph Partitioning and Sparse Matrix Ordering},
  booktitle = {Proceedings of the 1998 ACM/IEEE Conference on Supercomputing (SC98)},
  year      = {1998},
  publisher = {IEEE Computer Society},
  address   = {Washington, DC, USA},
  doi       = {10.1109/SC.1998.10018}
}

@ARTICLE{Sadowski2017,
       author = {{S{\c{a}}dowski}, Aleksander and {Wielgus}, Maciek and {Narayan}, Ramesh and {Abarca}, David and {McKinney}, Jonathan C. and {Chael}, Andrew},
        title = "{Radiative, two-temperature simulations of low-luminosity black hole accretion flows in general relativity}",
      journal = {\mnras},
         year = 2017,
        month = apr,
       volume = {466},
       number = {1},
        pages = {705-725},
          doi = {10.1093/mnras/stw3116},
archivePrefix = {arXiv},
       eprint = {1605.03184},
 primaryClass = {astro-ph.HE},
       adsurl = {https://ui.adsabs.harvard.edu/abs/2017MNRAS.466..705S}
}

@ARTICLE{Porth2026,
       author = {{Porth}, Oliver and {Kelly}, Adrian and {Willocx}, Olaf and {Wu}, Hao and {Vos}, Jesse and {Zhou}, Yuhao and {Olivares S{\'a}nchez}, H{\'e}ctor R. and {Oostrum}, Leon and {Hidding}, Johan and {Azizi}, Victor and {Xia}, Chun and {Keppens}, Rony and {Teunissen}, Jannis},
        title = "{Astrophysics on GPUs: introducing AGILE 1.0}",
      journal = {arXiv e-prints},
         year = 2026,
        month = jul,
          eid = {arXiv:2607.19277},
        pages = {arXiv:2607.19277},
          doi = {10.48550/arXiv.2607.19277},
archivePrefix = {arXiv},
       eprint = {2607.19277},
 primaryClass = {astro-ph.IM},
       adsurl = {https://ui.adsabs.harvard.edu/abs/2026arXiv260719277P}
}

@ARTICLE{Stone2026,
       author = {{Stone}, James M. and {Mullen}, Patrick D. and {Fielding}, Drummond and {Grete}, Philipp and {Guo}, Minghao and {Kempski}, Philipp and {Most}, Elias R. and {White}, Christopher J. and {Wong}, George N.},
        title = "{AthenaK: A Performance-portable Version of the Athena++ Adaptive Mesh Refinement Framework}",
      journal = {\apjs},
         year = 2026,
        month = mar,
       volume = {283},
       number = {1},
          eid = {27},
        pages = {27},
          doi = {10.3847/1538-4365/ae3717},
archivePrefix = {arXiv},
       eprint = {2409.16053},
 primaryClass = {astro-ph.IM},
       adsurl = {https://ui.adsabs.harvard.edu/abs/2026ApJS..283...27S}
}

@ARTICLE{Keppens2023,
       author = {{Keppens}, R. and {Popescu Braileanu}, B. and {Zhou}, Y. and {Ruan}, W. and {Xia}, C. and {Guo}, Y. and {Claes}, N. and {Bacchini}, F.},
        title = "{MPI-AMRVAC 3.0: Updates to an open-source simulation framework}",
      journal = {\aap},
         year = 2023,
        month = may,
       volume = {673},
          eid = {A66},
        pages = {A66},
          doi = {10.1051/0004-6361/202245359},
archivePrefix = {arXiv},
       eprint = {2303.03026},
 primaryClass = {astro-ph.IM},
       adsurl = {https://ui.adsabs.harvard.edu/abs/2023A&A...673A..66K}
}

@ARTICLE{Liska2020,
       author = {{Liska}, M.~T.~P. and {Chatterjee}, K. and {Issa}, D. and {Yoon}, D. and {Kaaz}, N. and {Tchekhovskoy}, A. and {van Eijnatten}, D. and {Musoke}, G. and {Hesp}, C. and {Rohoza}, V. and {Markoff}, S. and {Ingram}, A. and {van der Klis}, M.},
        title = "{H-AMR: A New GPU-accelerated GRMHD Code for Exascale Computing with 3D Adaptive Mesh Refinement and Local Adaptive Time Stepping}",
      journal = {\apjs},
         year = 2022,
        month = dec,
       volume = {263},
       number = {2},
          eid = {26},
        pages = {26},
          doi = {10.3847/1538-4365/ac9966},
       adsurl = {https://ui.adsabs.harvard.edu/abs/2022ApJS..263...26L}
}

@ARTICLE{Mckinney2017,
       author = {{McKinney}, Jonathan C. and {Chluba}, Jens and {Wielgus}, Maciek and {Narayan}, Ramesh and {Sadowski}, Aleksander},
        title = "{Double Compton and Cyclo-Synchrotron in Super-Eddington Discs, Magnetized Coronae, and Jets}",
      journal = {\mnras},
         year = 2017,
        month = may,
       volume = {467},
       number = {2},
        pages = {2241-2265},
          doi = {10.1093/mnras/stx227},
archivePrefix = {arXiv},
       eprint = {1608.08627},
 primaryClass = {astro-ph.HE},
       adsurl = {https://ui.adsabs.harvard.edu/abs/2017MNRAS.467.2241M}
}

@ARTICLE{Migone2007,
       author = {{Mignone}, A. and {Bodo}, G. and {Massaglia}, S. and {Matsakos}, T. and {Tesileanu}, O. and {Zanni}, C. and {Ferrari}, A.},
        title = "{PLUTO: A Numerical Code for Computational Astrophysics}",
      journal = {\apjs},
         year = 2007,
        month = may,
       volume = {170},
       number = {1},
        pages = {228-242},
          doi = {10.1086/513316},
archivePrefix = {arXiv},
       eprint = {astro-ph/0701854},
 primaryClass = {astro-ph},
       adsurl = {https://ui.adsabs.harvard.edu/abs/2007ApJS..170..228M}
}

@ARTICLE{Etienne2015,
       author = {{Etienne}, Zachariah B. and {Paschalidis}, Vasileios and {Haas}, Roland and {M{\"o}sta}, Philipp and {Shapiro}, Stuart L.},
        title = "{IllinoisGRMHD: an open-source, user-friendly GRMHD code for dynamical spacetimes}",
      journal = {Classical and Quantum Gravity},
         year = 2015,
        month = sep,
       volume = {32},
       number = {17},
          eid = {175009},
        pages = {175009},
          doi = {10.1088/0264-9381/32/17/175009},
archivePrefix = {arXiv},
       eprint = {1501.07276},
 primaryClass = {astro-ph.HE},
       adsurl = {https://ui.adsabs.harvard.edu/abs/2015CQGra..32q5009E}
}

@ARTICLE{Liska2024,
       author = {{Liska}, M.~T.~P. and {Kaaz}, N. and {Chatterjee}, K. and {Emami}, Razieh and {Musoke}, G.},
        title = "{Magnetic Flux Plays an Important Role during a Black Hole X-Ray Binary Outburst in Radiative Two-temperature General Relativistic Magnetohydrodynamic Simulations}",
      journal = {\apj},
         year = 2024,
        month = may,
       volume = {966},
       number = {1},
          eid = {47},
        pages = {47},
          doi = {10.3847/1538-4357/ad344a},
archivePrefix = {arXiv},
       eprint = {2309.15926},
 primaryClass = {astro-ph.HE},
       adsurl = {https://ui.adsabs.harvard.edu/abs/2024ApJ...966...47L}
}

@ARTICLE{Anninos2005,
       author = {{Anninos}, Peter and {Fragile}, P. Chris and {Salmonson}, Jay D.},
        title = "{Cosmos++: Relativistic Magnetohydrodynamics on Unstructured Grids with Local Adaptive Refinement}",
      journal = {\apj},
         year = 2005,
        month = dec,
       volume = {635},
       number = {1},
        pages = {723-740},
          doi = {10.1086/497294},
archivePrefix = {arXiv},
       eprint = {astro-ph/0509254},
 primaryClass = {astro-ph},
       adsurl = {https://ui.adsabs.harvard.edu/abs/2005ApJ...635..723A}
}

@article{Liska2018A,
author = {{Liska}, M. and {Hesp}, C. and {Tchekhovskoy}, A. and {Ingram}, A. and {van der Klis}, M. and {Markoff}, S.},
title = "{Formation of precessing jets by tilted black hole discs in 3D general relativistic MHD simulations}",
journal = {\mnras},
volume = {474},
number = {1},
pages = {L81-L85},
year = {2018},
month = {feb},
doi = {10.1093/mnrasl/slx174},
URL = {http://dx.doi.org/10.1093/mnrasl/slx174},
eprint = {/oup/backfile/content_public/journal/mnrasl/474/1/10.1093_mnrasl_slx174/4/slx174.pdf}
}

@ARTICLE{Bacchini2019,
       author = {{Bacchini}, F. and {Ripperda}, B. and {Porth}, O. and {Sironi}, L.},
        title = "{Generalized, Energy-conserving Numerical Simulations of Particles in General Relativity. II. Test Particles in Electromagnetic Fields and GRMHD}",
      journal = {\apjs},
         year = 2019,
        month = feb,
       volume = {240},
       number = {2},
          eid = {40},
        pages = {40},
          doi = {10.3847/1538-4365/aafcb3},
archivePrefix = {arXiv},
       eprint = {1810.00842},
 primaryClass = {astro-ph.HE},
       adsurl = {https://ui.adsabs.harvard.edu/abs/2019ApJS..240...40B}
}

@ARTICLE{Courant1928,
   author = {{Courant}, R. and {Friedrichs}, K. and {Lewy}, H.},
    title = "{{\"U}ber die partiellen Differenzengleichungen der mathematischen Physik}",
  journal = {Mathematische Annalen},
     year = 1928,
   volume = 100,
    pages = {32-74},
      doi = {10.1007/BF01448839},
   adsurl = {https://ui.adsabs.harvard.edu/abs/1928MatAn.100...32C}
}

@ARTICLE{Berger1989,
   author = {{Berger}, M.~J. and {Colella}, P.},
    title = "{Local adaptive mesh refinement for shock hydrodynamics}",
  journal = {Journal of Computational Physics},
     year = 1989,
    month = may,
   volume = 82,
    pages = {64-84},
      doi = {10.1016/0021-9991(89)90035-1},
   adsurl = {https://ui.adsabs.harvard.edu/abs/1989JCoPh..82...64B}
}

@ARTICLE{Balsara2001,
       author = {{Balsara}, Dinshaw S.},
        title = "{Divergence-Free Adaptive Mesh Refinement for Magnetohydrodynamics}",
      journal = {Journal of Computational Physics},
         year = "2001",
        month = "Dec",
       volume = {174},
       number = {2},
        pages = {614-648},
          doi = {10.1006/jcph.2001.6917},
archivePrefix = {arXiv},
       eprint = {astro-ph/0112150},
 primaryClass = {astro-ph},
       adsurl = {https://ui.adsabs.harvard.edu/abs/2001JCoPh.174..614B}
}

@ARTICLE{Zhang2026,
       author = {{Zhang}, Lizhong and {Stone}, James M. and {Davis}, Shane W. and {Jiang}, Yan-Fei and {Mullen}, Patrick D. and {White}, Christopher J.},
        title = "{Radiation GRMHD Models of Accretion onto Stellar-Mass Black Holes: III. Near-Eddington Accretion}",
      journal = {arXiv e-prints},
         year = 2026,
        month = mar,
          eid = {arXiv:2603.05588},
        pages = {arXiv:2603.05588},
          doi = {10.48550/arXiv.2603.05588},
archivePrefix = {arXiv},
       eprint = {2603.05588},
 primaryClass = {astro-ph.HE},
       adsurl = {https://ui.adsabs.harvard.edu/abs/2026arXiv260305588Z}
}

@ARTICLE{Zhang2025,
       author = {{Zhang}, Lizhong and {Stone}, James M. and {White}, Christopher J. and {Davis}, Shane W. and {Jiang}, Yan-Fei and {Mullen}, Patrick D.},
        title = "{Radiation GRMHD Models of Accretion onto Stellar-mass Black Holes. II. Super-Eddington Accretion}",
      journal = {\apj},
         year = 2026,
        month = apr,
       volume = {1001},
       number = {2},
          eid = {138},
        pages = {138},
          doi = {10.3847/1538-4357/ae5521},
archivePrefix = {arXiv},
       eprint = {2509.10638},
 primaryClass = {astro-ph.HE},
       adsurl = {https://ui.adsabs.harvard.edu/abs/2026ApJ..1001..138Z}
}

@ARTICLE{Zhang2024,
       author = {{Zhang}, Lizhong and {Stone}, James M. and {Mullen}, Patrick D. and {Davis}, Shane W. and {Jiang}, Yan-Fei and {White}, Christopher J.},
        title = "{Radiation GRMHD Models of Accretion onto Stellar-mass Black Holes. I. Survey of Eddington Ratios}",
      journal = {\apj},
         year = 2025,
        month = dec,
       volume = {995},
       number = {1},
          eid = {26},
        pages = {26},
          doi = {10.3847/1538-4357/ae0f91},
archivePrefix = {arXiv},
       eprint = {2506.02289},
 primaryClass = {astro-ph.HE},
       adsurl = {https://ui.adsabs.harvard.edu/abs/2025ApJ...995...26Z}
}

@ARTICLE{Trent2024,
       author = {{Trent}, Tyler and {Roley}, Karin and {Golden}, Matthew and {Psaltis}, Dimitrios and {{\"O}zel}, Feryal},
        title = "{Covariant Guiding Center Equations for Charged Particle Motions in General Relativistic Spacetimes}",
      journal = {\apj},
         year = 2025,
        month = jul,
       volume = {987},
       number = {1},
          eid = {101},
        pages = {101},
          doi = {10.3847/1538-4357/adbca8},
archivePrefix = {arXiv},
       eprint = {2404.01391},
 primaryClass = {astro-ph.HE},
       adsurl = {https://ui.adsabs.harvard.edu/abs/2025ApJ...987..101T}
}

@ARTICLE{Trent2025,
       author = {{Trent}, Tyler and {Psaltis}, Dimitrios and {{\"O}zel}, Feryal},
        title = "{A Hybrid Algorithm for Drift-Kinetic Particle Dynamics within General Relativistic Magnetohydrodynamics Simulations of Black Holes Accretion Flows}",
      journal = {arXiv e-prints},
         year = 2025,
        month = jul,
          eid = {arXiv:2507.11616},
        pages = {arXiv:2507.11616},
          doi = {10.48550/arXiv.2507.11616},
archivePrefix = {arXiv},
       eprint = {2507.11616},
 primaryClass = {astro-ph.HE},
       adsurl = {https://ui.adsabs.harvard.edu/abs/2025arXiv250711616T}
}

@ARTICLE{Ryan2015,
       author = {{Ryan}, B.~R. and {Dolence}, J.~C. and {Gammie}, C.~F.},
        title = "{bhlight: General Relativistic Radiation Magnetohydrodynamics with Monte Carlo Transport}",
      journal = {\apj},
         year = 2015,
        month = jul,
       volume = {807},
       number = {1},
          eid = {31},
        pages = {31},
          doi = {10.1088/0004-637X/807/1/31},
archivePrefix = {arXiv},
       eprint = {1505.05119},
 primaryClass = {astro-ph.HE},
       adsurl = {https://ui.adsabs.harvard.edu/abs/2015ApJ...807...31R}
}

@ARTICLE{Fragile2026,
       author = {{Fragile}, P. Chris and {Middleton}, Matthew J. and {Brasseur}, Brooks and {Bollimpalli}, Deepika A. and {Smith}, Zach},
        title = "{The nature of tilted supercritical accretion discs}",
      journal = {\mnras},
         year = 2026,
        month = may,
       volume = {548},
       number = {3},
          eid = {stag711},
        pages = {stag711},
          doi = {10.1093/mnras/stag711},
archivePrefix = {arXiv},
       eprint = {2604.11794},
 primaryClass = {astro-ph.HE},
       adsurl = {https://ui.adsabs.harvard.edu/abs/2026MNRAS.548ag711F}
}

@article{fry00,
	Adsurl = {http://adsabs.harvard.edu/cgi-bin/nph-bib_query?bibcode=2000ApJS..131..273F&db_key=AST},
	Author = {{Fryxell}, B. and {Olson}, K. and {Ricker}, P. and {Timmes}, F.~X. and {Zingale}, M. and {Lamb}, D.~Q. and {MacNeice}, P. and {Rosner}, R. and {Truran}, J.~W. and {Tufo}, H.},
	Doi = {10.1086/317361},
	Journal = {\apjs},
	Month = nov,
	Pages = {273-334},
	Title = {{FLASH: An Adaptive Mesh Hydrodynamics Code for Modeling Astrophysical Thermonuclear Flashes}},
	Volume = 131,
	Year = 2000}

@article{Gammie2003,
	Adsurl = {http://adsabs.harvard.edu/cgi-bin/nph-bib_query?bibcode=2003ApJ...589..444G&db_key=AST},
	Author = {{Gammie}, C.~F. and {McKinney}, J.~C. and {T{\'o}th}, G.},
	Doi = {10.1086/374594},
	Eprint = {astro-ph/0301509},
	Journal = {\apj},
	Month = may,
	Pages = {444-457},
	Title = {{HARM: A Numerical Scheme for General Relativistic Magnetohydrodynamics}},
	Volume = 589,
	Year = 2003}

@ARTICLE{Kaaz2022,
       author = {{Kaaz}, Nicholas and {Liska}, Matthew T.~P. and {Jacquemin-Ide}, Jonatan and {Andalman}, Zachary L. and {Musoke}, Gibwa and {Tchekhovskoy}, Alexander and {Porth}, Oliver},
        title = "{Nozzle Shocks, Disk Tearing and Streamers Drive Rapid Accretion in 3D GRMHD Simulations of Warped Thin Disks}",
      journal = {arXiv e-prints},
         year = 2022,
        month = oct,
          eid = {arXiv:2210.10053},
        pages = {arXiv:2210.10053},
archivePrefix = {arXiv},
       eprint = {2210.10053},
 primaryClass = {astro-ph.HE},
       adsurl = {https://ui.adsabs.harvard.edu/abs/2022arXiv221010053K}
}

@ARTICLE{Liska2022,
       author = {{Liska}, M.~T.~P. and {Musoke}, G. and {Tchekhovskoy}, A. and {Porth}, O. and {Beloborodov}, A.~M.},
        title = "{Formation of Magnetically Truncated Accretion Disks in 3D Radiation-transport Two-temperature GRMHD Simulations}",
      journal = {\apjl},
         year = 2022,
        month = aug,
       volume = {935},
       number = {1},
          eid = {L1},
        pages = {L1},
          doi = {10.3847/2041-8213/ac84db},
archivePrefix = {arXiv},
       eprint = {2201.03526},
 primaryClass = {astro-ph.HE},
       adsurl = {https://ui.adsabs.harvard.edu/abs/2022ApJ...935L...1L}
}

@ARTICLE{Musoke2022,
       author = {{Musoke}, G. and {Liska}, M. and {Porth}, O. and {van der Klis}, M. and {Ingram}, A.},
        title = "{Disk Tearing Leads to Low and High Frequency Quasi Periodic Oscillations in a GRMHD Simulation of a Thin Accretion Disk}",
      journal = {arXiv e-prints},
         year = 2022,
        month = jan,
          eid = {arXiv:2201.03085},
        pages = {arXiv:2201.03085},
archivePrefix = {arXiv},
       eprint = {2201.03085},
 primaryClass = {astro-ph.HE},
       adsurl = {https://ui.adsabs.harvard.edu/abs/2022arXiv220103085M}
}

@ARTICLE{Liska2023,
       author = {{Liska}, M.~T.~P. and {Kaaz}, N. and {Musoke}, G. and {Tchekhovskoy}, A. and {Porth}, O.},
        title = "{Radiation Transport Two-temperature GRMHD Simulations of Warped Accretion Disks}",
      journal = {\apjl},
         year = 2023,
        month = feb,
       volume = {944},
       number = {2},
          eid = {L48},
        pages = {L48},
          doi = {10.3847/2041-8213/acb6f4},
archivePrefix = {arXiv},
       eprint = {2210.10198},
 primaryClass = {astro-ph.HE},
       adsurl = {https://ui.adsabs.harvard.edu/abs/2023ApJ...944L..48L}
}

@Article{Porth2017,
  author        = {{Porth}, O. and {Olivares}, H. and {Mizuno}, Y. and {Younsi}, Z. and {Rezzolla}, L. and {Moscibrodzka}, M. and {Falcke}, H. and {Kramer}, M.},
  title         = {{The black hole accretion code}},
  journal       = {Computational Astrophysics and Cosmology},
  year          = {2017},
  volume        = {4},
  month         = may,
  adsurl        = {http://adsabs.harvard.edu/abs/2017ComAC...4....1P},
  doi           = {10.1186/s40668-017-0020-2},
  eprint        = {1611.09720},
  primaryclass  = {gr-qc},
}

@ARTICLE{Scepi2023,
       author = {{Scepi}, Nicolas and {Begelman}, Mitchell C. and {Dexter}, Jason},
        title = "{Magnetic support, wind-driven accretion, coronal heating, and fast outflows in a thin magnetically arrested disc}",
      journal = {arXiv e-prints},
         year = 2023,
        month = feb,
          eid = {arXiv:2302.10226},
        pages = {arXiv:2302.10226},
          doi = {10.48550/arXiv.2302.10226},
archivePrefix = {arXiv},
       eprint = {2302.10226},
 primaryClass = {astro-ph.HE},
       adsurl = {https://ui.adsabs.harvard.edu/abs/2023arXiv230210226S}
}

@Article{Foucart2018,
  author        = {{Foucart}, F.},
  title         = {{Monte Carlo closure for moment-based transport schemes in general relativistic radiation hydrodynamic simulations}},
  journal       = {\mnras},
  year          = {2018},
  volume        = {475},
  pages         = {4186-4207},
  month         = apr,
  __markedentry = {[Matthew:6]},
  adsnote       = {Provided by the SAO/NASA Astrophysics Data System},
  adsurl        = {http://adsabs.harvard.edu/abs/2018MNRAS.475.4186F},
  archiveprefix = {arXiv},
  doi           = {10.1093/mnras/sty108},
  eprint        = {1708.08452},
  keywords      = {gravitation, radiative transfer, methods: numerical, stars: black holes, stars: neutron},
  primaryclass  = {astro-ph.HE},
}

@ARTICLE{Sadowski2013,
       author = {{S{\k{a}}dowski}, Aleksander and {Narayan}, Ramesh and {Tchekhovskoy}, Alexander and {Zhu}, Yucong},
        title = "{Semi-implicit scheme for treating radiation under M1 closure in general relativistic conservative fluid dynamics codes}",
      journal = {\mnras},
         year = 2013,
        month = mar,
       volume = {429},
       number = {4},
        pages = {3533-3550},
          doi = {10.1093/mnras/sts632},
archivePrefix = {arXiv},
       eprint = {1212.5050},
 primaryClass = {astro-ph.HE},
       adsurl = {https://ui.adsabs.harvard.edu/abs/2013MNRAS.429.3533S}
}
\bibliographystyle{aasjournalv7.1}


\end{document}